\documentclass[12pt]{article}
\usepackage{graphicx}
\usepackage{amsfonts}
\usepackage{amssymb}
\usepackage{mathrsfs}
\usepackage{amsmath}
\usepackage[unicode]{hyperref}
\usepackage[numbers,sort&compress]{natbib}

\begin{document}
\renewcommand{\thefootnote}{\fnsymbol{footnote}}
\begin{titlepage}

\vspace{10mm}
\begin{center}
{\Large\bf Massive vector particles tunneling from black holes influenced by the generalized uncertainty principle}
\vspace{8mm}

{{\large Xiang-Qian Li${}^{1}$\footnote{E-mail address: lixiangqian13b@mails.ucas.ac.cn}}\\

\vspace{6mm}
${}^{1}${\normalsize \em School of Physics, University of Chinese Academy of Sciences, Beijing 100049, China}

}
\end{center}

\vspace{10mm}
\centerline{{\bf{Abstract}}}
\vspace{6mm}
%\noindent
This study considers the generalized uncertainty principle, which incorporates the central idea of large extra dimensions, to investigate the processes involved when massive spin-1 particles tunnel from Reissner-Nordstrom and Kerr black holes under the effects of quantum gravity. For the black hole, the quantum gravity correction decelerates the increase in temperature. Up to $\mathcal{O}(\frac{1}{M_f^2})$, the corrected temperatures are affected by the mass and angular momentum of the emitted vector bosons. In addition, the temperature of the Kerr black hole becomes uneven due to rotation. When the mass of the black hole approaches the order of the higher dimensional Planck mass $M_f$, it stops radiating and yields a black hole remnant.
\vskip 20pt

\noindent

\end{titlepage}
\newpage
\renewcommand{\thefootnote}{\arabic{footnote}}
\setcounter{footnote}{0}
\setcounter{page}{2}

\section{Introduction}
\label{Introduction}
Hawking stated that black holes can release radiation thermodynamically due to quantum vacuum fluctuation effects near the event horizon \cite{hawking}. Subsequently, Hawking radiation has attracted much attention from theoretical physicists and various methods have been proposed for deriving Hawking radiation. In particular, a semiclassical derivation was developed that models Hawking radiation as a tunneling process, which includes the null geodesic method and Hamilton-Jacobi method. The null geodesic method was first proposed by Kraus and Wilczek~\cite{pkfw1,pkfw2}, and then developed further by Parikh and Wilczek~\cite{mkpf,mkp1,mkp2}. The Hamilton-Jacobi method was proposed by Angheben \emph{et al}.~\cite{mamn} as an extension of Padmanabhan's methods~\cite{kstp,sskt}. Both approaches to tunneling rely on the fact that the tunneling probability for the classically forbidden trajectory from inside to outside the horizon is given by $\Gamma={\rm exp}\left(-2{\rm Im}I/\hbar\right)$, where $I$ is the classical action of the trajectory. These two methods differ in how the imaginary part of the classical action is calculated. Many useful results have been obtained using the null geodesic and Hamilton-Jacobi methods~\cite{ecv,ajmm,maam,zzhao1,zzhao2,rkrbm,ajmme,pmit,dysz,kern1,kern2,Chen:2011mg,Kruglov:2014iya,sik,grch1,grch2,grch3,xqli,Sakalli1,Sakalli2,Sakalli3}.

A common feature of various quantum gravity theories, such as string theory, loop quantum gravity, and noncommutative geometry, is the existence of a minimum measurable length~\cite{kppplb,maggi,garayIJMP,amelinoIJMP}. The generalized uncertainty principle (GUP) is a simple way of realizing this minimal length~\cite{Kempfprd,Scardigli:1999jh,Scardigli:2003kr}. An effective model of the GUP in one-dimensional quantum mechanics, which incorporates the central idea of large extra dimensions, was given by~\cite{Hossenfelder:2003jz}
\begin{eqnarray}
\label{GUP1}L_f k(p) &=& \tanh \left( \frac{p}{M_f}
\right), \\
\label{GUP2}L_f \omega(E) &=& \tanh \left( \frac{E}{M_f} \right),
\end{eqnarray}
where the generators of the translations in space and time are the wave vector $k$ and the frequency $\omega$, and $L_f$ and $M_f$ are the higher dimensional minimal length and Planck mass, respectively. $L_f$ and $M_f$ satisfy $L_fM_f =\hbar$. The quantization in position
representation $\hat{x}=x$ leads to
\begin{equation}\label{komega}
k= - {\mathrm i} \partial_x, \omega=  + {\mathrm i} \partial_t.
\end{equation}
Therefore, the low energy limit $p\ll M_f$ including the order of $(p/{M_f})^3$ gives
\begin{eqnarray}
\label{modifiedp}p&\approx& -i\hbar\partial_x\left(1-\beta\hbar^2\partial_x^2\right),\\
\label{modifiedE}E&\approx& i\hbar\partial_t\left(1-\beta\hbar^2\partial_t^2\right),
\end{eqnarray}
where $\beta=1/(3M_f^2)$. Then, the modified commutation relation is given by
\begin{equation}\label{commutation}
\left[x,p\right]= i \hbar \left(1+ \beta p^2\right),
\end{equation}
and the generalized uncertainty relation (GUR) is
\begin{equation}\label{GUP}
\Delta x \Delta p \geq \frac{\hbar}{2}\left[1+ \beta \langle p^2\rangle\right].
\end{equation}
From Eqs~\eqref{commutation} and~\eqref{GUP}, it can be concluded that the departure of GUP from the Heisenberg uncertainty principle increases with the momentum of the particle. We note that Eqs~\eqref{modifiedp}--\eqref{GUP} only apply to particles in the low energy limit $p\ll M_f$, which is the specific case considered in the present study. In the low energy regime, the parameter $\beta$ should be constrained in experiments designed to test the uncertainty principle, such as those by~\cite{nairzpra,Pikovski:2011zk}. Other generalized uncertainty relations can be found in previous studies. A widely discussed relation, $\Delta x \Delta p \geq \frac{\hbar}{2}\left[1+ l^2\frac{\Delta p^2}{\hbar^2}\right]$, was proposed based on some aspects of quantum gravity and string theory~\cite{Kempfprd}, where the cutoff $l$ was selected as a string scale in the context of the perturbative string theory or Plank scale based on quantum gravity. Another interesting GUR was obtained by treating the mass source as a Gaussian wave function and the horizon as a horizon wave function~\cite{Casadio:2015rwa}, i.e., $\Delta r \simeq l_p\frac{m_p}{\Delta p}+\gamma l_p \frac{\Delta p}{m_p}$, where the first part represents the uncertainty of the radial size of the source and the second represents the horizon uncertainty, and $\gamma$ is a parameter that represents the order of unity in the full quantum gravity regime, which becomes very small in the semiclassical regime.

Black holes are an important research area in the study of quantum gravity effects and many studies of black hole physics have incorporated the GUP. The thermodynamics of black holes have been investigated in the framework of GUP~\cite{aliJHEP,majumPLB,binaPRD,chenNPPS,adlerGRG,xiangwenJHEP,kimsonJHEP}. By combining the GUP with the tunneling method, Nozari and Mehdipour studied the modified tunneling rate of a Schwarzschild black hole~\cite{nozariEPL}. The GUP-deformed
Hamilton-Jacobi equation for fermions in curved spacetime was introduced and the corrected Hawking temperatures were derived for various types of spacetime in~\cite{majumGRG,chenahep,chenjhep,chenjcap,barguenoPLB,mubenrong,pengwang,anaclePLB,fengEPJC}. By studying the tunneling of fermions, it was found that the quantum gravity effects slowed down the increase in the Hawking temperatures, where this property naturally leads to a residual mass during black hole evaporation.

In this study, we investigate massive spin-1 particles ($W^{\pm}$, $Z^0$) tunneling across the horizons of black
holes using the Hamilton-Jacobi method, which incorporates the minimal length effect
via Eqs~\eqref{modifiedp} and~\eqref{modifiedE}. Our calculations show that the quantum gravity correction is
related to the black hole's mass as well as to the mass and angular momentum of the
emitted vector bosons. Furthermore, the quantum gravity correction explicitly retards the increase in temperature during the
black hole evaporation process. As a result, the quantum correction will balance the traditional tendency for a temperature increase at some point during the evaporation, which leads to the existence of remnants.

The remainder of this paper is organized as follows. In Section~\ref{Section2}, based on the GUP-corrected Lagrangian of the massive vector field, we derive the equation of motion for the vector bosons in curved spacetime. In Section~\ref{Section3}, by incorporating
GUP, we investigate the tunneling of charged massive bosons in a Reissner-Nordstrom
black hole. The tunneling of massive bosons in a Kerr black hole is also studied and the
remnants are derived in Section~\ref{Section4}. Section~\ref{Section5} provides some discussion and the conclusions of this study. We use the spacelike metric signature convention $(-,+,+,+)$ in this study.

\section{Generalized field equations for massive vector bosons}
\label{Section2}
We start from the kinetic term of the uncharged vector boson field in flat spacetime within the framework of GUP, $\frac{1}{2}\widetilde{\mathfrak{B}}_{\mu\nu}\widetilde{\mathfrak{B}}^{\mu\nu}$, where the modified field strength tensor is given by
\begin{equation}\label{mfs}
\widetilde{\mathfrak{B}}_{\mu\nu}=\left(1-\beta\hbar^2\partial_{\mu}^2\right)\partial_{\mu}\mathfrak{B}_{\nu}-\left(1-\beta\hbar^2\partial_{\nu}^2\right)\partial_{\nu}\mathfrak{B}_{\mu}.
\end{equation}
It should be noted that additional derivative terms exist. Next, we generalize this to the case of a charged vector boson field ($W^{\pm}$) in charged black hole spacetime. Considering the gauge principle, the additional derivatives also act on the local unitary transformation operator $U(x)$, so they must also be replaced by covariant derivatives~\cite{Kober:2010sj}:
\begin{eqnarray}
\label{tfl}\left(1-\beta\hbar^2\partial_{0}^2\right)\partial_{0} &\rightarrow& \left(1+\beta\hbar^2g^{00}{D^{\pm}_{0}}^2\right)D^{\pm}_{0}, \\
\label{xfl}\left(1-\beta\hbar^2\partial_{i}^2\right)\partial_{i} &\rightarrow& \left(1-\beta\hbar^2g^{ii}{D^{\pm}_{i}}^2\right)D^{\pm}_{i},
\end{eqnarray}
where $D^{\pm}_{\mu}=\nabla_{\mu}\pm\frac{i}{\hbar}eA_{\mu}$ with $\nabla_{\mu}$ is the geometrically covariant derivative, $A_{\mu}$ is the electromagnetic field of the black hole, and $e$ denotes the charge of the $W^{+}$ boson. The difference in signs of the $\mathcal{O(\beta)}$ terms in Eqs~\eqref{tfl} and~\eqref{xfl} is attributable to the fact that $g^{00}$ always shares different signs with $g^{ii}$.

By defining $\mathcal{D}^{\pm}_{0}=\left(1+\beta\hbar^2g^{00}{D^{\pm}_{0}}^2\right)D^{\pm}_{0}$ and $\mathcal{D}^{\pm}_{i}=\left(1-\beta\hbar^2g^{ii}{D^{\pm}_{i}}^2\right)D^{\pm}_{i}$, the GUP-corrected Lagrangian of $W$-boson field is given by
\begin{equation}\label{Lgupcurved}
\mathcal{L}^{GUP}= -\frac{1}{2}\left(\mathcal{D}^{+}_{\mu}W^{+}_{\nu}-\mathcal{D}^{+}_{\nu}W^{+}_{\mu}\right)\left(\mathcal{D}^{-\mu}W^{-\nu}-\mathcal{D}^{-\nu}W^{-\mu}\right)-\frac{m_{W}^2}{{\hbar}^2}W^{+}_{\mu}W^{-\mu}
-\frac{i}{\hbar}eF^{\mu\nu}W^{+}_{\mu}W^{-}_{\nu},
\end{equation}
where $F_{\mu\nu}=\widehat{\nabla}_{\mu}A_{\nu}-\widehat{\nabla}_{\nu}A_{\mu}$, with $\widehat{\nabla}_{0}=\left(1+\beta{\hbar}^2g^{00}{\nabla_{0}}^2\right)\nabla_{0}$ and $\widehat{\nabla}_{i}=\left(1-\beta{\hbar}^2g^{ii}{\nabla_{i}}^2\right)\nabla_{i}$.
Accordingly, the corresponding generalized action should be
\begin{equation}\label{Gupaction}
\mathcal{S}^{GUP}=\int \text{d}x^{4}\sqrt{-g}\mathcal{L}^{GUP}\left(W^{\pm}_{\mu},\partial_{\mu}W^{\pm}_{\nu},\partial_{\mu}\partial_{\rho}W^{\pm}_{\nu},\partial_{\mu}\partial_{\rho}\partial_{\lambda}W^{\pm}_{\nu}\right).
\end{equation}
This action is invariant under a local $U\left(1\right)$ gauge transformation, which does not refer to spacetime transformation.

By varying the action~\eqref{Gupaction} with respect to the fields $W^{-}$ and $W^{+}$, it follows immediately that
\begin{eqnarray}\label{Fe0}
% \nonumber to remove numbering (before each equation)
\frac{\partial \mathcal{S}}{\partial W^{-}_{\nu}}-\partial_{\mu}\frac{\partial \mathcal{S}}{\partial \left(\partial_{\mu}W^{-}_{\nu}\right)}+\partial_{\mu}\partial_{\rho}\frac{\partial \mathcal{S}}{\partial \left(\partial_{\mu}\partial_{\rho}W^{-}_{\nu}\right)}-\partial_{\mu}\partial_{\rho}\partial_{\lambda}\frac{\partial \mathcal{S}}{\partial \left(\partial_{\mu}\partial_{\rho}\partial_{\lambda}W^{-}_{\nu}\right)} &=&0, \\
\frac{\partial \mathcal{S}}{\partial W^{+}_{\nu}}-\partial_{\mu}\frac{\partial \mathcal{S}}{\partial \left(\partial_{\mu}W^{+}_{\nu}\right)}+\partial_{\mu}\partial_{\rho}\frac{\partial \mathcal{S}}{\partial \left(\partial_{\mu}\partial_{\rho}W^{+}_{\nu}\right)}-\partial_{\mu}\partial_{\rho}\partial_{\lambda}\frac{\partial \mathcal{S}}{\partial \left(\partial_{\mu}\partial_{\rho}\partial_{\lambda}W^{+}_{\nu}\right)} &=&0.
\end{eqnarray}
Then, by substituting the GUP Lagrangian~\eqref{Lgupcurved} in~\eqref{Fe0}, we obtain
\begin{eqnarray} \label{Fe1}
&&\partial_{\mu}\left(\sqrt{-g}W^{+\mu\nu}\right)-3\beta\partial_{0}\left[\sqrt{-g}g^{00}\left(e^2{A_{0}}^2+i\hbar e\nabla_{0}A_{0}\right)W^{+0\nu}\right]                                        \notag\\
&&+3\beta\partial_{i}\left[\sqrt{-g}g^{ii}\left(e^2{A_{i}}^2+i\hbar e\nabla_{i}A_{i}\right)W^{+i\nu}\right]+3\beta\partial_{0}\partial_{0}\left(\sqrt{-g}g^{00}i\hbar eA_0W^{+0\nu}\right)  \notag\\
&&-3\beta\partial_{i}\partial_{i}\left(\sqrt{-g}g^{ii}i\hbar eA_iW^{+i\nu}\right)+\beta\hbar^2\partial_{0}\partial_{0}\partial_{0}\left(\sqrt{-g}g^{00}W^{+0\nu}\right)    \notag\\
&&-\beta\hbar^2\partial_{i}\partial_{i}\partial_{i}\left(\sqrt{-g}g^{ii}W^{+i\nu}\right)+\sqrt{-g}\frac{i}{\hbar}e A_{\mu}W^{+\mu\nu}-\sqrt{-g}\frac{m_W^2}{\hbar^2}W^{+\nu}  \notag\\
&&-\sqrt{-g}\frac{i}{\hbar}eF^{\mu\nu}W_{\mu}^{+}+\beta \sqrt{-g}g^{00}\left(i\hbar e\nabla_{0}\nabla_{0}A_0+3e^2A_0\nabla_0A_0-\frac{i}{\hbar}e^3{A_0}^3\right)W^{+0\nu}  \notag\\
&&-\beta \sqrt{-g}g^{ii}\left(i\hbar e\nabla_{i}\nabla_{i}A_i+3e^2A_i\nabla_iA_i-\frac{i}{\hbar}e^3{A_i}^3\right)W^{+i\nu}=0,
\end{eqnarray}
where we denote $\mathcal{D}^{+}_{\mu}W^{+}_{\nu}-\mathcal{D}^{+}_{\nu}W^{+}_{\mu}$ by $W^{+}_{\mu\nu}$. This is the equation of motion for the $W^{+}$ boson field and we can repeat the same procedure to obtain the equation of motion for the $W^{-}$ boson field. By setting $e$ to $0$, we obtain the field equation for massive bosons in the case of uncharged bosons or in an uncharged spacetime background
\begin{equation}\label{Fe2}
\partial_{\mu}\left(\sqrt{-g}\mathfrak{B}^{\mu\nu}\right)-\sqrt{-g}\frac{m^2}{\hbar^2}\mathfrak{B}^{\nu}+\beta\hbar^2\partial_{0}\partial_{0}\partial_{0}\left(\sqrt{-g}g^{00}\mathfrak{B}^{0\nu}\right)-\beta\hbar^2\partial_{i}\partial_{i}\partial_{i}\left(\sqrt{-g}g^{ii}\mathfrak{B}^{i\nu}\right)=0.
\end{equation}

\section{Massive vector particles tunneling from a Reissner-Nordstrom black hole}
\label{Section3}
The Reissner-Nordstrom black hole describes a spherically symmetric static spacetime with
charge $Q$. The metric is given by
\begin{equation}\label{RNmetric}
 ds^2=-f(r)dt^2+g(r)^{-1}dr^2+r^2\left(d\theta^2+\text{sin}^2\theta d\phi^2\right),
\end{equation}
with the electromagnetic potential
\begin{equation}\label{Empotential}
 A_{\mu}=\left(A_0,0,0,0\right)=\left(\frac{Q}{r},0,0,0\right),
\end{equation}
where
\begin{equation}\label{RNmetricfu}
 f(r)=g(r)=1-\frac{2M}{r}+\frac{Q^2}{r^2}=\frac{(r-r_{+})(r-r_{-})}{r^2},
\end{equation}
and $r_{\pm}=M\pm\sqrt{M^2-Q^2}$ represents the locations of the outer horizon and the inner horizon, respectively. In this study, without any loss of generality, we only consider the tunneling process for $W^{+}$ bosons. The calculation is similar
for the $W^{-}$ case.

According to the WKB approximation, $W^{+}_\mu$ has the form of
\begin{equation} \label{WKBW}
W^{+}_\mu=C_\mu(t,r,\theta,\phi) {\rm exp}\left[\frac{i}{\hbar}S(t,r,\theta,\phi)\right],
\end{equation}
where $S$ is defined as
\begin{equation} \label{S0123}
    S(t,r,\theta,\phi)=S_0(t,r,\theta,\phi)+\hbar S_1(t,r,\theta,\phi)+\hbar^2 S_2(t,r,\theta,\phi)+\cdots.
\end{equation}
By substituting Eqs~\eqref{WKBW},~\eqref{S0123}, and the Reissner-Nordstrom metric~\eqref{RNmetric} into Equation~\eqref{Fe1}, and keeping only the lowest order in $\hbar$, we obtain the equations for the coefficients $C_\mu$
\begin{eqnarray}
\label{Wproca1}g(r)\left[C_0(\partial_rS_0)^2\mathcal{P}_{1}^{2}-C_1(\partial_rS_0)(\partial_tS_0+eA_{t})\mathcal{P}_{1}\mathcal{P}_{0}\right] &&\notag\\
+\frac{1}{r^2}\left[C_0(\partial_{\theta}S_0)^2\mathcal{P}_{2}^{2}-C_2(\partial_{\theta}S_0)(\partial_tS_0+eA_{t})\mathcal{P}_{2}\mathcal{P}_{0}\right] &&\notag\\
+\frac{1}{r^2{\rm sin}^2\theta}\left[C_0(\partial_{\phi}S_0)^2\mathcal{P}_{3}^{2}-C_3(\partial_{\phi}S_0)(\partial_tS_0+eA_{t})\mathcal{P}_{3}\mathcal{P}_{0}\right]+C_0m_{W}^2= 0,&&
\end{eqnarray}
\begin{eqnarray}
\label{Wproca2}-\frac{1}{f(r)}\left[C_1(\partial_tS_0+eA_{t})^2\mathcal{P}_{0}^{2}-C_0(\partial_tS_0+eA_{t})(\partial_rS_0)\mathcal{P}_{0}\mathcal{P}_{1}\right] &&\notag\\
+\frac{1}{r^2}\left[C_1(\partial_{\theta}S_0)^2\mathcal{P}_{2}^{2}-C_2(\partial_{\theta}S_0)(\partial_rS_0)\mathcal{P}_{2}\mathcal{P}_{1}\right] &&\notag\\
+\frac{1}{r^2{\rm sin}^2\theta}\left[C_1(\partial_{\phi}S_0)^2\mathcal{P}_{3}^{2}-C_3(\partial_{\phi}S_0)(\partial_rS_0)\mathcal{P}_{3}\mathcal{P}_{1}\right]+C_1m_{W}^2= 0,&&
\end{eqnarray}
\begin{eqnarray}
\label{Wproca3}-\frac{1}{f(r)}\left[C_2(\partial_tS_0+eA_{t})^2\mathcal{P}_{0}^{2}-C_0(\partial_tS_0+eA_{t})(\partial_{\theta}S_0)\mathcal{P}_{0}\mathcal{P}_{2}\right] && \notag\\
+g(r)\left[C_2(\partial_rS_0)^2\mathcal{P}_{1}^{2}-C_1(\partial_rS_0)(\partial_{\theta}S_0)\mathcal{P}_{1}\mathcal{P}_{2}\right] && \notag\\
+\frac{1}{r^2{\rm sin}^2\theta}\left[C_2(\partial_{\phi}S_0)^2\mathcal{P}_{3}^{2}-C_3(\partial_{\phi}S_0)(\partial_{\theta}S_0)\mathcal{P}_{3}\mathcal{P}_{2}\right]+C_2m_{W}^2= 0, &&
\end{eqnarray}
\begin{eqnarray}
\label{Wproca4}-\frac{1}{f(r)}\left[C_3(\partial_tS_0+eA_{t})^2\mathcal{P}_{0}^{2}-C_0(\partial_tS_0+eA_{t})(\partial_{\phi}S_0)\mathcal{P}_{0}\mathcal{P}_{3}\right] && \notag\\
+g(r)\left[C_3(\partial_rS_0)^2\mathcal{P}_{1}^{2}-C_1(\partial_rS_0)(\partial_{\phi}S_0)\mathcal{P}_{1}\mathcal{P}_{3}\right] && \notag\\
+\frac{1}{r^2}\left[C_3(\partial_{\theta}S_0)^2\mathcal{P}_{2}^{2}-C_2(\partial_{\theta}S_0)(\partial_{\phi}S_0)\mathcal{P}_{2}\mathcal{P}_{3}\right]+C_3m_{W}^2= 0, &&
\end{eqnarray}
where the $\mathcal{P}_{\mu}$s are defined as
\begin{eqnarray}
&&\mathcal{P}_{0}=1+\beta\frac{1}{f(r)}(\partial_tS_0+eA_{t})^2,\ \mathcal{P}_{1}=1+\beta g(r)(\partial_rS_0)^2, \notag\\
&&\mathcal{P}_{2}=1+\beta\frac{1}{r^2}(\partial_{\theta}S_0)^2,\ \mathcal{P}_{3}=1+\beta \frac{1}{r^2{\rm sin}^2\theta}(\partial_{\phi}S_0)^2.
\end{eqnarray}
Considering the property of Reissner-Nordstrom spacetime and the question that we aim to address, then following the standard process, we separate the variables
\begin{equation}\label{RNs0fj}
 S_0 = -Et+W(r)+\Theta(\theta,\phi),
\end{equation}
where $E$ is the energy of the emitted vector particles. By inserting Eq.~\eqref{RNs0fj} into Eqs~\eqref{Wproca1}--\eqref{Wproca4}, we can obtain a matrix equation
\begin{equation}\label{matrixeq}
K(C_{0}, C_{1}, C_{2}, C_{3})^{T}=0,
\end{equation}
where $K$ is a 4$\times$4 matrix, the elements of which are
\begin{eqnarray}
&&K_{11}= g(r)W'^2\mathcal{P}_{1}^{2}+\frac{{J_{\theta}}^2}{r^2}\mathcal{P}_{2}^{2}+\frac{{J_{\phi}}^2}{r^2{\rm sin}^2\theta}\mathcal{P}_{3}^{2}+m_{W}^2,\ K_{12}=-g(r)W'(-E+eA_{t})\mathcal{P}_{1}\mathcal{P}_{0}, \notag\\
&&K_{13}=-\frac{J_{\theta}(-E+eA_{t})}{r^2}\mathcal{P}_{2}\mathcal{P}_{0},\ K_{14}=-\frac{J_{\phi}(-E+eA_{t})}{r^2{\rm sin}^2\theta}\mathcal{P}_{3}\mathcal{P}_{0},             \notag\\
&&K_{21}=\frac{(-E+eA_t)W'}{f(r)}\mathcal{P}_{0}\mathcal{P}_{1},\ K_{22}=-\frac{(-E+eA_t)^2}{f(r)}\mathcal{P}_{0}^{2}+\frac{{J_{\theta}}^2}{r^2}\mathcal{P}_{2}^{2}+\frac{J_{\phi}^2}{r^2{\rm sin}^2\theta}\mathcal{P}_{3}^{2}+m_{W}^2, \notag\\
&&K_{23}=-\frac{J_{\theta}W'}{r^2}\mathcal{P}_{2}\mathcal{P}_{1},\ K_{24}=-\frac{J_{\phi}W'}{r^2{\rm sin}^2\theta}\mathcal{P}_{3}\mathcal{P}_{1}, \notag\\
&&K_{31}=\frac{(-E+eA_t)J_{\theta}}{f(r)}\mathcal{P}_{0}\mathcal{P}_{2},\ K_{32}=-g(r)W'J_{\theta}\mathcal{P}_{1}\mathcal{P}_{2},       \\
&&K_{33}=-\frac{(-E+eA_t)^2}{f(r)}\mathcal{P}_{0}^{2}+g(r)W'^2\mathcal{P}_{1}^{2}+\frac{J_{\phi}^2}{r^2{\rm sin}^2\theta}\mathcal{P}_{3}^{2}+m_{W}^2,\ K_{34}=-\frac{J_{\theta}J_{\phi}}{r^2{\rm sin}^2\theta}\mathcal{P}_{3}\mathcal{P}_{2},        \notag\\
&&K_{41}=\frac{(-E+eA_t)J_{\phi}}{f(r)}\mathcal{P}_{0}\mathcal{P}_{3},\ K_{42}=-g(r)W'J_{\phi}\mathcal{P}_{1}\mathcal{P}_{3},        \notag\\
&&K_{43}=-\frac{J_{\theta}J_{\phi}}{r^2}\mathcal{P}_{2}\mathcal{P}_{3},\ K_{44}=-\frac{(-E+eA_t)^2}{f(r)}\mathcal{P}_{0}^{2}+g(r)W'^2\mathcal{P}_{1}^{2}+\frac{J_{\theta}^2}{r^2}\mathcal{P}_{2}^{2}+m_{W}^2,        \notag
\end{eqnarray}
where $W'=\partial_rW$, $J_{\theta}=\partial_{\theta}\Theta$ and $J_{\phi}=\partial_{\phi}\Theta$.

Eq.~\eqref{matrixeq} has a nontrivial solution if the determinant of the matrix $K$ equals zero. ${\rm det} K=0$ should yield the following equation
\begin{eqnarray}\label{Wequation}
\mathcal{O}(\beta^6)(\partial_rW)^{18}+\mathcal{O}(\beta^5)(\partial_rW)^{16}+\mathcal{O}(\beta^4)(\partial_rW)^{14}+\mathcal{O}(\beta^3)(\partial_rW)^{12} &&\notag\\
+\mathcal{O}(\beta^2)(\partial_rW)^{10}+\left[A_8+\mathcal{O}(\beta^2)\right](\partial_rW)^{8}+\left[A_6+\mathcal{O}(\beta^2)\right](\partial_rW)^{6} &&\notag\\
+\left[A_4+\mathcal{O}(\beta^2)\right](\partial_rW)^{4}+\left[A_2+\mathcal{O}(\beta^2)\right](\partial_rW)^{2}+A_0+\mathcal{O}(\beta^2)=0
\end{eqnarray}
(please refer to Appendix~\ref{appendix1} for the definition of $A_i$s). By neglecting the higher order terms of $\beta$ and solving Eq.~\eqref{Wequation}, we obtain the solution to the derivative of the radial action
\begin{equation}\label{wpr}
\partial_rW=\pm\sqrt{-\frac{m^2}{g(r)}+\frac{(E-eA_t)^2}{f(r)g(r)}-\frac{J^2_{\theta}+J^2_{\phi}{\rm csc}^2\theta}{g(r)r^2}}\left(1+\frac{\mathcal{X}_1}{\mathcal{X}_2}\beta\right),
\end{equation}
where
\begin{eqnarray}
\mathcal{X}_1&=&-3f(r)m^4r^2+6m^2r^2(E-eA_t)^2-6f(r)m^2(J^2_{\theta}+J^2_{\phi}{\rm csc}^2\theta)-\frac{6f(r)J^4_{\theta}}{r^2} \notag\\
&+&6(E-eA_t)^2(J^2_{\theta}+J^2_{\phi}{\rm csc}^2\theta)-\frac{7f(r)J^2_{\theta}J^2_{\phi}{\rm csc}^2\theta}{r^2}-\frac{3f(r)J^4_{\theta}J^2_{\phi}{\rm csc}^2\theta}{2m^2r^4} \notag\\
&-&\frac{5f(r)J^4_{\phi}{\rm csc}^4\theta}{r^2}+\frac{3f(r)J^2_{\theta}J^4_{\phi}{\rm csc}^4\theta}{2m^2r^4}, \\
\mathcal{X}_2&=&-f(r)m^2r^2+r^2(E-eA_t)^2-f(r)(J^2_{\theta}+J^2_{\phi}{\rm csc}^2\theta).
\end{eqnarray}

Integrating Eq.~\eqref{wpr} around the pole at the outer horizon $r_{+}=M+\sqrt{M^2-Q^2}$ yields the solution of the radial action. The particle's tunneling rate is determined by the imaginary part of the action,
\begin{eqnarray}\label{wrim}
Im W_\pm (r) & = &\pm Im \int dr \sqrt{-\frac{m^2}{g(r)}+\frac{(E-eA_t)^2}{f(r)g(r)}-\frac{J^2_{\theta}+J^2_{\phi}{\rm csc}^2\theta}{g(r)r^2}}\left(1+\frac{\mathcal{X}_1}{\mathcal{X}_2}\beta\right)\nonumber \\
& = & \pm \pi\frac{r_+^2}{r_+ - r_-} (E - eA_{t+})
\times\left(1+ \beta\Xi\right),
\end{eqnarray}
where $\Xi=6m^2+\frac{6}{r_+^2}\left(J^2_{\theta}+J^2_{\phi}{\rm csc}^2\theta\right)$. It is quite clear that $\Xi>0$. We note that $W_{+}$ represents the radial function for the outgoing particles and $W_{-}$ is for the ingoing particles. Thus, the tunneling rate of $W^{+}$ bosons at the outer event horizon is
\begin{eqnarray} \label{kerrtunnelproba}
\Gamma&=&\frac{P_{outgoing}}{P_{ingoing}}=\frac{{\rm exp}\left[-\frac{2}{\hbar}({\rm Im} W_{+}+{\rm Im} \Theta)\right]}{{\rm exp}\left[-\frac{2}{\hbar}({\rm Im} W_{-}+{\rm Im} \Theta)\right]}={\rm exp}\left[-\frac{4}{\hbar}{\rm Im} W_{+}\right] \notag\\
&=&\exp\left[ -\frac{4\pi}{\hbar}\frac{r_+^2}{r_+ - r_-} (E - eA_{t+})
\times\left(1 + \beta\Xi\right)\right].
\end{eqnarray}
If we set $\hbar=1$, then the effective Hawking temperature is deduced as
\begin{eqnarray}\label{effectiveHT}
T_{e-H}= \frac{r_+ - r_-}{4\pi r_+^2\left(1+
\beta\Xi\right)}= T_0\left(1 -\beta\Xi\right),
\end{eqnarray}
where $T_0 =  \frac{r_+ - r_-}{4\pi r_+^2}$ is the original Hawking temperature of a Reissner-Nordstrom black hole. From Eq.~\eqref{effectiveHT}, it can be inferred that the corrected temperature relies on the quantum numbers (mass and angular momentum) of the emitted vector bosons. Moreover, the quantum effects explicitly counteract the temperature increase during evaporation, which will cancel it out at some point. Naturally, black hole remnants will be left.

\section{Massive vector particles tunneling from a Kerr black hole}
\label{Section4}
In this section, we investigate the tunneling of massive vector particles at the outer event horizon of a Kerr black hole where we consider the GUP. For simplicity, we suppose that the emitted vector particles are uncharged, so the motion of the vector field is described by Eq.~\eqref{Fe2}.
The line element within Kerr spacetime is given by
\begin{eqnarray} \label{metric1}
ds^2&=&-(1-\frac{2Mr}{\rho^2})dt^2+\frac{\rho^2}{\Delta}dr^2+\rho^2d\theta^2\notag \\
&&+\left[(r^2+a^2)+\frac{2Mra^2{\rm sin}^2\theta}{\rho^2}\right]{\rm sin}^2\theta d\varphi^2-\frac{4Mra{\rm sin}^2\theta}{\rho^2}dtd\varphi,
\end{eqnarray}
where $\rho^2=r^2+a^2{\rm cos}^2\theta$, $\Delta=r^2-2Mr+a^2$, $M$ is the black hole mass, and $a$ is the angular momentum per unit mass. To ensure that the event horizon
coincides with the infinite red-shift surface, we introduce a new coordinate $\chi=\varphi-\Omega t$ with $\Omega=\frac{2Mra}{(r^2+a^2)^2-\Delta a^2{\rm sin}^2\theta}$, and thus the metric \eqref{metric1} becomes
\begin{equation} \label{metric2}
ds^2=-\frac{\Delta \rho^2}{\Sigma(r,\theta)}dt^2+\frac{\rho^2}{\Delta}dr^2+{\rho}^2d\theta^2+\frac{\Sigma(r,\theta)}{\rho^2}{\rm sin}^2\theta d\chi^2,
\end{equation}
where $\Sigma(r,\theta)=(r^2+a^2)^2-\Delta a^2{\rm sin}^2\theta$.

According to the WKB approximation, $\mathfrak{B}_\mu$ has the form of
\begin{equation} \label{WKB2}
\mathfrak{B}_\mu=C_\mu(t,r,\theta,\chi) {\rm exp}\left[\frac{i}{\hbar}S(t,r,\theta,\chi)\right],
\end{equation}
where $S$ is defined as
\begin{equation} \label{S01232}
    S(t,r,\theta,\chi)=S_0(t,r,\theta,\chi)+\hbar S_1(t,r,\theta,\chi)+\hbar^2 S_2(t,r,\theta,\chi)+\cdots.
\end{equation}

By substituting Eqs~\eqref{WKB2},~\eqref{S01232}, and the Kerr metric~\eqref{metric2} into~\eqref{Fe2}, and keeping only the lowest order in $\hbar$, we obtain the equations for the coefficients $C_\mu$
\begin{eqnarray}
\label{kerrceq1}&&\frac{\Delta}{\rho^2}\left[C_0(\partial_rS_0)^2\mathcal{P}_{1}^{2}-C_1(\partial_rS_0)(\partial_tS_0)\mathcal{P}_{1}\mathcal{P}_{0}\right]+\frac{1}{\rho^2}\left[C_0(\partial_{\theta}S_0)^2\mathcal{P}_{2}^{2}-C_2(\partial_{\theta}S_0)(\partial_tS_0)\mathcal{P}_{2}\mathcal{P}_{0}\right] \notag\\
&&+\frac{\rho^2}{\Sigma{\rm sin}^2\theta}\left[C_0(\partial_{\chi}S_0)^2\mathcal{P}_{3}^{2}-C_3(\partial_{\chi}S_0)(\partial_tS_0)\mathcal{P}_{3}\mathcal{P}_{0}\right]+C_0m^2= 0, \\
\label{kerrceq2}&&\frac{-\Sigma}{\Delta\rho^2}\left[C_1(\partial_tS_0)^2\mathcal{P}_{0}^{2}-C_0(\partial_tS_0)(\partial_rS_0)\mathcal{P}_{0}\mathcal{P}_{1}\right]+\frac{1}{\rho^2}\left[C_1(\partial_{\theta}S_0)^2\mathcal{P}_{2}^{2}-C_2(\partial_{\theta}S_0)(\partial_rS_0)\mathcal{P}_{2}\mathcal{P}_{1}\right] \notag\\
&&+\frac{\rho^2}{\Sigma{\rm sin}^2\theta}\left[C_1(\partial_{\chi}S_0)^2\mathcal{P}_{3}^{2}-C_3(\partial_{\chi}S_0)(\partial_rS_0)\mathcal{P}_{3}\mathcal{P}_{1}\right]+C_1m^2= 0, \\
\label{kerrceq3}&&\frac{-\Sigma}{\Delta\rho^2}\left[C_2(\partial_tS_0)^2\mathcal{P}_{0}^{2}-C_0(\partial_tS_0)(\partial_{\theta}S_0)\mathcal{P}_{0}\mathcal{P}_{2}\right]+\frac{\Delta}{\rho^2}\left[C_2(\partial_rS_0)^2\mathcal{P}_{1}^{2}-C_1(\partial_rS_0)(\partial_{\theta}S_0)\mathcal{P}_{1}\mathcal{P}_{2}\right] \notag\\
&&+\frac{\rho^2}{\Sigma{\rm sin}^2\theta}\left[C_2(\partial_{\chi}S_0)^2\mathcal{P}_{3}^{2}-C_3(\partial_{\chi}S_0)(\partial_{\theta}S_0)\mathcal{P}_{3}\mathcal{P}_{2}\right]+C_2m^2= 0, \\
\label{kerrceq4}&&\frac{-\Sigma}{\Delta\rho^2}\left[C_3(\partial_tS_0)^2\mathcal{P}_{0}^{2}-C_0(\partial_tS_0)(\partial_{\chi}S_0)\mathcal{P}_{0}\mathcal{P}_{3}\right]+\frac{\Delta}{\rho^2}\left[C_3(\partial_rS_0)^2\mathcal{P}_{1}^{2}-C_1(\partial_rS_0)(\partial_{\chi}S_0)\mathcal{P}_{1}\mathcal{P}_{3}\right] \notag\\
&&+\frac{1}{\rho^2}\left[C_3(\partial_{\theta}S_0)^2\mathcal{P}_{2}^{2}-C_2(\partial_{\theta}S_0)(\partial_{\chi}S_0)\mathcal{P}_{2}\mathcal{P}_{3}\right]+C_3m^2= 0,
\end{eqnarray}
where the $\mathcal{P}_{\mu}$s are defined as
\begin{eqnarray}
&&\mathcal{P}_{0}=1+\beta\frac{\Sigma}{\Delta\rho^2}(\partial_tS_0)^2,\ \mathcal{P}_{1}=1+\beta \frac{\Delta}{\rho^2}(\partial_rS_0)^2, \notag\\
&&\mathcal{P}_{2}=1+\beta\frac{1}{\rho^2}(\partial_{\theta}S_0)^2,\ \mathcal{P}_{3}=1+\beta \frac{\rho^2}{\Sigma{\rm sin}^2\theta}(\partial_{\chi}S_0)^2.
\end{eqnarray}

Considering the properties of Kerr spacetime, we separate the variables as
\begin{eqnarray}\label{kerrs0fj}
S_0 &&= -Et+W(r)+j\varphi+\Theta(\theta) \notag\\
&&= -(E-j\Omega)t+W(r)+j\chi+ \Theta(\theta),
\end{eqnarray}
where $E$ and $j$ denote the energy and angular momentum of the emitted particle, respectively. By inserting Eq.~\eqref{kerrs0fj} into Eqs~\eqref{kerrceq1}--\eqref{kerrceq4}, we can obtain a matrix equation $K\left(C_{0},C_{1},C_{2},C_{3}\right)^T=0$ and the elements of $K$ are expressed as
\begin{eqnarray}
&&K_{11}= \frac{\Delta W'^2}{\rho^2}\mathcal{P}_{1}^{2}+\frac{{J_{\theta}}^2}{\rho^2}\mathcal{P}_{2}^{2}+\frac{j^2\rho^2}{\Sigma {\rm sin}^2\theta}\mathcal{P}_{3}^{2}+m^2,\ K_{12}=\frac{\Delta W'(E-j\Omega)}{\rho^2}\mathcal{P}_{1}\mathcal{P}_{0},\notag\\
&&K_{13}=\frac{J_{\theta}(E-j\Omega)}{\rho^2}\mathcal{P}_{2}\mathcal{P}_{0},\ K_{14}=\frac{\rho^2j(E-j\Omega)}{\Sigma {\rm sin}^2\theta}\mathcal{P}_{3}\mathcal{P}_{0},\ K_{21}=\frac{-\Sigma W'(E-j\Omega)}{\Delta\rho^2}\mathcal{P}_{0}\mathcal{P}_{1}, \notag \\
&&K_{22}=\frac{\Sigma(E-j\Omega)^2}{\Delta\rho^2}\mathcal{P}_{0}^{2}+\frac{{J_{\theta}}^2}{\rho^2}\mathcal{P}_{2}^{2}+\frac{j^2\rho^2}{\Sigma {\rm sin}^2\theta}\mathcal{P}_{3}^{2}+m^2,\ K_{23}=\frac{-J_{\theta}W'}{\rho^2}\mathcal{P}_{2}\mathcal{P}_{1},\notag\\
&&K_{24}=\frac{-\rho^2jW'}{\Sigma{\rm sin}^2\theta}\mathcal{P}_{3}\mathcal{P}_{1},\ K_{31}=\frac{-\Sigma J_{\theta}(E-j\Omega)}{\Delta \rho^2}\mathcal{P}_{0}\mathcal{P}_{2},\ K_{32}=\frac{-\Delta J_{\theta}W'}{\rho^2}\mathcal{P}_{1}\mathcal{P}_{2}, \\
&&K_{33}=\frac{-\Sigma (E-j\Omega)^2}{\Delta \rho^2}\mathcal{P}_{0}^{2}+\frac{\Delta{W'}^2}{\rho^2}\mathcal{P}_{1}^{2}+\frac{\rho^2j^2}{\Sigma {\rm sin}^2\theta}\mathcal{P}_{3}^{2}+m^2,\ K_{34}=\frac{-\rho^2jJ_{\theta}}{\Sigma {\rm sin}^2\theta}\mathcal{P}_{3}\mathcal{P}_{2}, \notag\\
&&K_{41}=\frac{-\Sigma j(E-j\Omega)}{\Delta \rho^2}\mathcal{P}_{0}\mathcal{P}_{3},\ K_{42}=\frac{-\Delta jW'}{\rho^2}\mathcal{P}_{1}\mathcal{P}_{3},\ K_{43}=\frac{-J_{\theta}j}{\rho^2}\mathcal{P}_{2}\mathcal{P}_{3}, \notag\\
&&K_{44}=\frac{-\Sigma (E-j\Omega)^2}{\Delta \rho^2}\mathcal{P}_{0}^{2}+\frac{\Delta{W'}^2}{\rho^2}\mathcal{P}_{1}^{2}+\frac{{J_{\theta}}^2}{\rho^2}\mathcal{P}_{2}^{2}+m^2, \notag
\end{eqnarray}
where $J_{\theta}$ is identified as $\partial_{\theta}S_{0}$.

The determination of the coefficient matrix should be equal to zero to ensure that Eqs. \eqref{kerrceq1}--\eqref{kerrceq4} have a nontrivial solution. ${\rm det} K=0$ yields the following equation
\begin{eqnarray}\label{kerrWequation}
\mathcal{O}(\beta^6)(\partial_rW)^{18}+\mathcal{O}(\beta^5)(\partial_rW)^{16}+\mathcal{O}(\beta^4)(\partial_rW)^{14}+\mathcal{O}(\beta^3)(\partial_rW)^{12} &&\notag\\
+\mathcal{O}(\beta^2)(\partial_rW)^{10}+\left[B_8+\mathcal{O}(\beta^2)\right](\partial_rW)^{8}+\left[B_6+\mathcal{O}(\beta^2)\right](\partial_rW)^{6} &&\notag\\
+\left[B_4+\mathcal{O}(\beta^2)\right](\partial_rW)^{4}+\left[B_2+\mathcal{O}(\beta^2)\right](\partial_rW)^{2}+B_0+\mathcal{O}(\beta^2)=0
\end{eqnarray}
(please refer to Appendix~\ref{appendix1} for definitions of the $B_i$s). By neglecting the higher order terms of $\beta$ and solving Eq.~\eqref{kerrWequation}, we obtain the solution to the derivative of the radial action
\begin{equation}\label{kerrwpr}
\partial_rW=\pm\sqrt{-\frac{m^2\rho^2}{\Delta}+\frac{\Sigma(E-j\Omega)^2}{\Delta^2}-\frac{J^2_{\theta}}{\Delta}-\frac{\rho^4j^2{\rm csc}^2\theta}{\Delta\Sigma}}\left(1+\frac{\mathcal{Y}_1}{\mathcal{Y}_2}\beta\right),
\end{equation}
where
\begin{eqnarray}
\mathcal{Y}_1&=&-3m^4\Delta\rho^2\Sigma+6m^2\Sigma^2(E-j\Omega)^2-6m^2\Delta\Sigma J^2_{\theta}+\frac{6\Sigma^2(E-j\Omega)^2J^2_{\theta}}{\rho^2}-\frac{6\Delta\Sigma J^4_{\theta}}{\rho^2} \notag\\
&-&6m^2\Delta\rho^4 j^2{\rm csc}^2\theta+6\rho^2\Sigma(E-j\Omega)^2 j^2{\rm csc}^2\theta-7\Delta\rho^2J^2_{\theta}j^2{\rm csc}^2\theta-\frac{3\Delta J^4_{\theta}j^2{\rm csc}^2\theta}{2m^2}  \notag\\
&-&\frac{5\Delta\rho^6j^4{\rm csc}^4\theta}{\Sigma}+\frac{3\Delta\rho^4 J^2_{\theta}j^4{\rm csc}^4\theta}{2m^2\Sigma}, \\
\mathcal{Y}_2&=&-m^2\Delta\rho^2\Sigma+\Sigma^2(E-j\Omega)^2-\Delta\Sigma J^2_{\theta}-\Delta\rho^4j^2{\rm csc}^2\theta.
\end{eqnarray}

Integrating Eq.~\eqref{kerrwpr} around the pole at the outer horizon $r_+ = M +\sqrt{M^2-a^2}$ yields the solution for the radial action. The particle's tunneling rate is determined by the imaginary part of the action,
\begin{eqnarray}\label{wrim}
Im W_\pm (r) & = &\pm Im \int dr \sqrt{-\frac{m^2\rho^2}{\Delta}+\frac{\Sigma(E-j\Omega)^2}{\Delta^2}-\frac{J^2_{\theta}}{\Delta}-\frac{\rho^4j^2{\rm csc}^2\theta}{\Delta\Sigma}}\left(1+\frac{\mathcal{Y}_1}{\mathcal{Y}_2}\beta\right)\nonumber \\
& = & \pm \pi \left(E -j \Omega _+\right)\frac{r_+^2 +a^2}{r_+ - r_-}
 \left( 1+  \beta \Pi\right),
\end{eqnarray}
where $\Pi=6m^2+\frac{6}{r_+^2+a^2{\rm cos}^2\theta}\left(J^2_{\theta}+j^2{\rm csc}^2\theta\right)$. It is obvious that $\Pi>0$. The tunneling rate of the vector bosons at the outer event horizon is
\begin{eqnarray} \label{tunnelproba}
\Gamma&=&\frac{P_{outgoing}}{P_{ingoing}}=\frac{{\rm exp}\left[-\frac{2}{\hbar}({\rm Im} W_{+}+{\rm Im} \Theta)\right]}{{\rm exp}\left[-\frac{2}{\hbar}({\rm Im} W_{-}+{\rm Im} \Theta)\right]}={\rm exp}\left[-\frac{4}{\hbar}{\rm Im} W_{+}\right] \notag\\
&=&\exp\left[-\frac{4\pi}{\hbar}\frac{r_+^2
+a^2}{r_+ - r_-}\left(E -j \Omega _+\right)\times\left( 1+ \beta \Pi\right)\right].
\end{eqnarray}
If we set $\hbar=1$, then the effective Hawking temperature is deduced as
\begin{eqnarray}\label{kerreffectiveHT}
T_{e-H}= \frac{r_+ - r_-}{4\pi(r_+^2 +a^2)}\frac{1}{\left( 1+
\beta \Pi \right)} = T_0(1-  \beta \Pi),
\end{eqnarray}
where $T_0 = \frac{r_+ - r_-}{4\pi(r_+^2 +a^2)} $ is the original Hawking temperature of a Kerr black hole. Similar to the results
obtained for a Reissner-Nordstrom black hole, the corrected temperature is lower than
the original Hawking temperature, where it is related to the
black hole's mass and angular momentum, as well as to the mass and angular momentum of the emitted vector bosons. It should be noted that due to the quantum gravity effect, the corrected Hawking temperature of a Kerr black hole become uneven since $\Pi$ is a function of $\theta$.

Dimensional reduction near the horizon can be used to study the standard processes of particle tunneling~\cite{Umetsu1,Umetsu2}, which is attributable to the fact that all large non-extremal black holes basically resemble Rindler space. For the standard Hawing radiation, all species of particles located very close to the horizon are effectively massless when considering infinite blueshift, so the Hawing temperatures of all particles are the same. However, our calculations also show that quantum gravity effects should make particles with different identities or quantum numbers differ in terms of their effective Hawking temperatures. When the particles approach the horizon, they can never be infinite-blueshifted because of the existence of minimal length.

When the quantum gravity effects are neglected, i.e., $\beta=0$, the standard Hawking temperatures of
Reissner-Nordstrom and Kerr black holes can be recovered by Eqs~\eqref{effectiveHT} and~\eqref{kerreffectiveHT}, respectively. To estimate the residual masses of the black holes at the level of the order of the magnitude, we consider the charge-free ($Q = 0$) and non-rotating ($a=0$) case, i.e., Schwarzschild spacetime, as a special case of both Reissner-Nordstrom and Kerr black holes. Then, the corrected Hawking
temperature
\begin{eqnarray}
T_{e-H}= \frac{1}{8\pi M} \left[1-6\beta
\left(m^{2}+\frac{J^2_{\theta}+J^2_{\phi}{\rm csc}^2\theta}{r_h^2}\right)\right] \label{eq5.2}
\end{eqnarray}
represents that of a Schwarzschild black hole. It should be noted that $\frac{J^2_{\theta}+J^2_{\phi}{\rm csc}^2\theta}{r_h^2}$ represents the kinetic energy component along the tangent plane of the horizon surface at the emission point. To estimate the order of magnitude of the residual mass, it is reasonable to approximate $m^{2}+\frac{J^2_{\theta}+J^2_{\phi}{\rm csc}^2\theta}{r_h^2}$ as $E^2$. To avoid the temperature $T_{e-H}$ becoming negative, the value of $E$ should satisfy $E<M_f$, where a factor of $1/\sqrt{2}$ is omitted. The temperature stops increasing when
\begin{equation}
\frac{1}{8\pi\left(M-dM\right)}-\frac{1}{8\pi M}\simeq\frac{\beta E^2}{8\pi M},
\end{equation}
which indicates that the temperature variation of the black hole caused by emitting a vector particle with energy $dM$ reaches equilibrium with that caused by the quantum gravity effect. Then, by using the condition that $dM = E$ and $\beta =1/M_f^2$, we obtain
\begin{equation}
M_{\hbox{Res}} \simeq \frac{M_f^2}{E} \gtrsim
M_f, \hspace{7mm} T_{\hbox{Res}} \lesssim
\frac{1}{8\pi M_f},
\label{final results}
\end{equation}
where an approximation of $E(M-E)$ to $EM$ is employed. $M_f$ is the higher dimensional Planck mass, which is related to the four-dimensional Planck mass $M_p$ by~\cite{Hossenfelder:2003jz}
\begin{equation}
M_p^2=R^dM_f^{d+2},
\end{equation}
where $R$ represents the compactification radius. The current lower limits on $M_f$ range from 3.67 Tev/$c^2$ for $d=2$ to 2.25 Tev/$c^2$ for $d=6$~\cite{Chatrchyan:2012me}.

Our calculations show that because of the minimal length effect, the black hole will stop emitting vector massive particles at some point, thereby leading to the residual mass of black hole evaporation. This conclusion is consistent with that obtained by~\cite{chenjcap,mubenrong} who studied the tunneling process for fermions.

\section{Discussion and conclusion}
\label{Section5}
In this study, we investigated the GUP effect on tunneling by massive vector particles from Reissner-Nordstrom and Kerr black holes. First, we derived a modified equation of motion for the massive vector bosons by generalizing the Lagrangian density of uncharged vector fields in flat spacetime within the Heisenberg uncertainty principle to that of charged vector fields in curved spacetime within the GUP. Using the WKB approximation and Hamilton-Jacobi ansatz, we derived the effective Hawking temperature of the black holes. Our results showed that if the effect of quantum gravity
is considered, then the behavior of a tunneling particle at the event horizon will differ from the original case, where the GUP-corrected temperature is highly dependent on the mass $M$, the electric charge $Q$, and the angular momentum $a$ of the black hole, as well as on the mass and angular momentum of the emitted vector particles. Up to $\mathcal{O}(\frac{1}{M_f^2})$, the effective Hawking temperature does not depend on the electric charge of the vector bosons. In addition, we found that the GUP-corrected Hawking temperature is smaller than the original case, where it stops increasing when the mass of the black hole reaches the minimal value $M_{Res}$, which is in the order of the higher dimensional Planck mass $M_f$.

In this study, we employed the GUP framework given by Eqs~\eqref{GUP1} and~\eqref{GUP2}, but several alternative forms of the GUP can be employed to study the tunneling of particles from a black hole horizon. For example, $p_i=p_{0i}(1+\frac{\beta_0}{M_p^2}p^2)$ was employed by~\cite{chenahep,chenjhep,chenjcap,pengwang,fengEPJC}. Different GUP forms may yield different results and further research is needed to clarify this issue.

\appendix
\section{Coefficients of the differential equations}
\label{appendix1}
In the following, we give the coefficients in Eq.~\eqref{Wequation}:
\begin{equation}\label{A8}
 A_8=6 m^2 \beta  g^4,
\end{equation}
\begin{equation}\label{A6}
 A_6=\frac{g^3 m^2}{fr^2}\Big\{r^2\big[f+12fm^2\beta-12(E-e A_t)^2\beta\big]+12f\beta J_{\theta}^2+12f\beta\text{csc}^2\theta J_{\phi}^2\Big\},
\end{equation}
\begin{eqnarray}\label{A4}
A_4&&=\frac{g^2}{fr^6}\bigg\{3f\beta J_{\theta}^4(4m^2r^2+\text{csc}^2\theta J_{\phi}^2)+J_{\theta}^2\Big[3m^2r^4\big(f+4m^2\beta-4(E-e A_t)^2\beta   \notag \\
&&+14fm^2r^2\beta\text{csc}^2\theta J_{\phi}^2)-3f\beta\text{csc}^4\theta J_{\phi}^4\Big]+m^2r^2\Big[3r^4\big(-(E-eA_t)^2(1+4m^2\beta)+f(m^2  \notag \\
&&+2m^4\beta)\big)+3r^2\big(f+4fm^2\beta-4(E-eA_t)^2\beta\big)\text{csc}^2\theta J_{\phi}^2+10f\beta\text{csc}^4\theta J_{\phi}^4\Big]\bigg\},
\end{eqnarray}

\begin{eqnarray}\label{A2}
A_2 &&=\frac{g\text{csc}^6\theta}{r^3r^8}\Big[r^2\big(fm^2-(E-e A_t)^2\big)\text{sin}^2\theta+f\text{sin}^2\theta J_{\theta}^2+f J_{\phi }^2 \Big]\bigg\{6f^2\beta\text{sin}^2\theta J_{\theta}^4(2m^2r^2\text{sin}^2\theta  \notag \\
&&+J_{\phi }^2)+f^2J_{\theta}^2(3m^2r^4\text{sin}^4\theta+4m^2r^2\beta\text{sin}^2\theta J_{\phi}^2-6\beta J_{\phi}^4)+m^2r^2\Big[3r^4\big(f^2m^2-f(E-e A_t)^2 \notag \\
&&-4(E-e A_t)^4\beta\big)\text{sin}^4\theta+3f^2r^2\text{sin}^2\theta J_{\phi}^2+8f^2\beta J_{\phi}^4\Big]\bigg\},
\end{eqnarray}

\begin{eqnarray}\label{A0}
% \nonumber to remove numbering (before each equation)
A_0 &&=\frac{1}{f^4 r^{10}}\Big\{r^2\big[f m^2-(E-e A_t)^2\big]+f J_{\theta}^2+f \text{csc}^2\theta J_{\phi
}^2\Big\}^2\bigg\{3 f^2 \beta  J_{\theta}^4 (2 m^2 r^2+\text{csc}^2\theta J_{\phi}^2)      \notag \\
&&+f^2 J_{\theta}^2 (m^2 r^4+2 m^2 r^2 \beta  \text{csc}^2\theta J_{\phi}^2-3 \beta  \text{csc}^4\theta J_{\phi}^4)+m^2 r^2 \Big[r^4 \big(f^2 m^2-f (E-e A_t)^2  \notag \\
&&-6 \beta(E-e A_t)^4\big)+f^2 r^2 \text{csc}^2\theta J_{\phi}^2+4 f^2 \beta  \text{csc}^4\theta J_{\phi}^4\Big]\bigg\},
\end{eqnarray}
where the arguments of $f(r)$ and $g(r)$ have been omitted.

Next, we give the coefficients in Eq.~\eqref{kerrWequation}:
\begin{equation}\label{B8}
B_8=\frac{6m^2\rho\Delta^4}{\rho^8},
\end{equation}

\begin{eqnarray}\label{B6}
B_6&&=\frac{m^2\Delta^2\text{csc}^2\theta}{\rho^{10}\Sigma}\bigg\{\rho^2\Big[12j^2\beta\Delta\rho^4+\Sigma\big(\Delta\rho^2+12m^2\beta\Delta\rho^2-12(E-j\Omega)^2\beta\Sigma\big)\text{sin}^2\theta\Big] \notag \\
&&+12\beta\Delta\rho^2\Sigma\text{sin}^2\theta J_{\theta}^2\bigg\},
\end{eqnarray}

\begin{eqnarray}\label{B4}
B_4&&=\frac{\Delta\text{csc}^4\theta}{\rho^{10}\Sigma^2}\bigg\{10j^4\beta\Delta\rho^{10}m^2+3j^2\rho^6\Sigma m^2\text{sin}^2\theta\big[(1+4m^2\beta)\Delta\rho^2-4(E-j\Omega)^2\beta\Sigma\big] \notag \\
&&+3\Sigma^2m^2\rho^4\text{sin}^4\theta\Big[2m^4\beta\Delta\rho^2-\Sigma(E-j\Omega)^2+m^2\big(\Delta\rho^2-4(E-j\Omega)^2\beta\Sigma\big)\Big]+\rho^2J_{\theta}^2 \notag \\
&&\Big[-3j^4\Delta\beta\rho^6+14j^2m^2\beta\Delta\rho^4\Sigma\text{sin}^2\theta+3m^2\Sigma^2\text{sin}^4\theta\big(\Delta\rho^2+4m^2\beta\Delta\rho^2\notag \\
&&-4(E-j\Omega)^2\beta\Sigma\big)\Big]+3\beta\Delta\rho^2\Sigma\text{sin}^2\theta J_{\theta}^4(j^2\rho^2+4m^2\Sigma\text{sin}^2\theta) \bigg\},
\end{eqnarray}

\begin{eqnarray}\label{B2}
B_2&&=\frac{\text{cos}^6\theta}{\rho^{14}\Sigma^3}\bigg\{\rho^2\Big[j^2\Delta\rho^4+\Sigma\text{sin}^2\theta\big(m^2\Delta\rho^2-(E-j\Omega)^2\Sigma\big)\Big]+\Delta\rho^2\Sigma\text{sin}^2\theta J_{\theta}^2\bigg\}\bigg\{m^2\rho^4 \notag \\
&&\Big[8j^4\beta\Delta\rho^8+3j^2\Delta\rho^6\Sigma\text{sin}^2\theta+3m^2\Delta\rho^4\Sigma^2\text{sin}^4\theta-3(E-j\Omega)^2\Sigma^3\text{sin}^4\theta\big(\Delta\rho^2 \notag \\
&&+4\beta\Sigma(E-j\Omega)^2\big)\Big]+\Delta\rho^6J_{\theta}^2(-6j^4\beta\rho^4+4j^2m^2\beta\rho^2\Sigma\text{sin}^2\theta+3m^2\Sigma^2\text{sin}^4\theta) \notag \\
&&+6\beta\Delta\rho^4\Sigma J_{\theta}^4\text{sin}^2\theta(j^2\rho^2+2m^2\Sigma\text{sin}^2\theta)\bigg\},
\end{eqnarray}

\begin{eqnarray}\label{B0}
B_0&&=\frac{\text{csc}^8\theta}{\Delta\rho^{16}\Sigma^4}\Big\{j^2\Delta\rho^6+\Sigma\rho^2\text{sin}^2\theta\big[m^2\Delta\rho^2-\Sigma(E-j\Omega)^2\big]+\Delta\rho^2\Sigma\text{sin}^2\theta J_{\theta}^2\Big\}^2\bigg\{m^2\rho^4    \notag \\
&&\Big[4j^4\beta\Delta\rho^8+j^2\Delta\rho^6\Sigma\text{sin}^2\theta+m^2\Delta\rho^4\Sigma^2\text{sin}^4\theta-(E-j\Omega)^2\Sigma^3\text{sin}^4\theta\big(\Delta\rho^2 \notag \\
&&+6\beta\Sigma(E-j\Omega)^2\big)\Big]+\Delta\rho^6J_{\theta}^2(-3j^4\beta\rho^4+2j^2m^2\beta\rho^2\Sigma\text{sin}^2\theta+m^2\Sigma^2\text{sin}^4\theta)
\notag \\
&&+3\beta\Delta\rho^4\Sigma J_{\theta}^4\text{sin}^2\theta(j^2\rho^2+2m^2\Sigma\text{sin}^2\theta)\bigg\}.
\end{eqnarray}

\section*{Acknowledgments}
This study was supported partly by the National Science Foundation of China under Grant No. 11475237, No. 11121064, and No. 10821504.


\begin{thebibliography}{99}
\bibitem{hawking}
S.W. Hawking, Commun. Math. Phys. 43, 199 (1975).

\bibitem{pkfw1}
P. Kraus, F. Wilczek, Mod. Phys. Lett. A 9, 3713 (1994).

\bibitem{pkfw2}
P. Kraus, F. Wilczek, Nucl. Phys. B 437, 231 (1995).

\bibitem{mkpf}
M.K. Parikh, F. Wilczek, Phys. Rev. Lett. 85, 5042 (2000).

\bibitem{mkp1}
M.K. Parikh, Phys. Lett. B 546, 189 (2002).

\bibitem{mkp2}
M.K. Parikh, Int. J. Mod. Phys. D 13, 2351 (2004).

\bibitem{mamn}
M. Angheben, M. Nadalini, L. Vanzo, S. Zerbini, J. High Energy Phys. 05,
014 (2005).

\bibitem{kstp}
K. Srinivasan, T. Padmanabhan, Phys. Rev. D 60, 024007 (1999).

\bibitem{sskt}
S. Shankaranarayanan, K. Srinivasan, T. Padmanabhan, Mod. Phys. Letts. 16, 571 (2001).

\bibitem{ecv}
E.C. Vagenas, Phys. Lett. B 503, 399 (2001).

\bibitem{ajmm}
A.J.M. Medved, Phys. Rev. D 66, 124009 (2002).

\bibitem{maam}
M. Arzano, A.J.M. Medved, E.C. Vagenas, J. High Energy Phys. 0509, 037 (2005).

\bibitem{zzhao1}
J.Y. Zhang, Z. Zhao, Phys. Lett. B 618, 14 (2005).

\bibitem{zzhao2}
J.Y. Zhang, Z. Zhao, J. High Energy Phys. 0505, 10055 (2005).

\bibitem{rkrbm}
R. Kerner, R.B. Mann, Phys. Rev. D 73, 104010 (2006).

\bibitem{ajmme}
A.J.M. Medved, E.C. Vagenas, Mod. Phys. Lett. A 20, 2499 (2005).

\bibitem{pmit}
P. Mitra, Phys. Lett. B 648, 240 (2007).

\bibitem{dysz}
D.Y. Chen, S.Z. Yang, Gen. Relativ. Gravitation 39, 1503 (2007).

\bibitem{kern1}
R. Kerner, R.B. Mann, Class. Quant. Grav. 25, 095014 (2008).

\bibitem{kern2}
R. Kerner, R.B. Mann, Phys. Lett. B 665, 277-283 (2008).

\bibitem{Chen:2011mg}
  D.~Chen, B.~Mu, H.~Wu, H.~Yang,
  %``One-dimensional quantum channel and Hawking radiation in the Kerr and Kerr-Newman black holes,''
  Int.\ J.\ Theor.\ Phys.\ 52, 1593 (2013).

\bibitem{Kruglov:2014iya}
S.~I.~Kruglov,
  %``Black hole emission of vector particles in (1+1) dimensions,''
  Int.\ J.\ Mod.\ Phys.\ A 29, 1450118 (2014).
  %[arXiv:1408.6561 [gr-qc]].
  %%CITATION = ARXIV:1408.6561;%%
  %6 citations counted in INSPIRE as of 28 juin 2015
\bibitem{sik}
S.I. Kruglov, Mod. Phys. Lett. A 29, 1450203 (2014).
\bibitem{grch1}
G.R. Chen, S. Zhou, Y.C. Huang, Int. J. Mod. Phys. D 24, 1550005 (2015).
\bibitem{grch2}
G.R. Chen, S. Zhou, Y.C. Huang, Astrophys. Space Sci. 357, 51 (2015).
\bibitem{grch3}
G.R. Chen, Y.C. Huang, Int. J. Mod. Phys. A 30, 1550083 (2015).
\bibitem{xqli}
X.~Q.~Li, G.~R.~Chen, Phys.\ Lett.\ B 751, 34 (2015).
\bibitem{Sakalli1}
  I.~Sakalli and A.~Ovgun,
  %``Quantum Tunneling of Massive Spin-1 Particles From Non-stationary Metrics,''
  Gen.\ Rel.\ Grav.\ 48, no. 1, 1 (2016).
\bibitem{Sakalli2}
  H.~Gursel, I.~Sakalli,
  %``Hawking Radiation of Massive Vector Particles From Warped AdS$_3$ Black Hole,''
  Can.\ J.\ Phys.\ 94, no. 2, 147 (2016).
\bibitem{Sakalli3}
  I.~Sakalli, A.~Ovgun,
  %``Tunnelling of vector particles from Lorentzian wormholes in 3+1 dimensions,''
  Eur.\ Phys.\ J.\ Plus 130, no. 6, 110 (2015).


\bibitem{kppplb}
K. Konishi, G. Paffuti, P. Provero, Phys. Lett. B 234, 276 (1990).
\bibitem{maggi}
M. Maggiore, Phys. Lett. B 319, 83 (1993).
\bibitem{garayIJMP}
L.J. Garay, Int. J. Mod. Phys. A 10, 145 (1995).
\bibitem{amelinoIJMP}
G. Amelino-Camelia, Int. J. Mod. Phys. D 11, 35 (2002).
\bibitem{Kempfprd}
A. Kempf, G. Mangano, R. B. Mann, Phys. Rev. D 52, 1108 (1995).

\bibitem{Scardigli:1999jh}
F.~Scardigli, Phys.\ Lett.\ B 452, 39 (1999).

\bibitem{Scardigli:2003kr}
F.~Scardigli, R.~Casadio, Class.\ Quant.\ Grav.\ 20, 3915 (2003).


\bibitem{Hossenfelder:2003jz}
  S.~Hossenfelder, M.~Bleicher, S.~Hofmann, J.~Ruppert, S.~Scherer, H.~Stoecker,
  Phys.\ Lett.\ B 575, 85 (2003).

\bibitem{nairzpra}
O. Nairz, M. Arndt, A. Zeilinger, Phys. Rev. A 65, 032109 (2002).

\bibitem{Pikovski:2011zk}
  I.~Pikovski, M.~R.~Vanner, M.~Aspelmeyer, M.~S.~Kim, C.~Brukner,
  %``Probing Planck-scale physics with quantum optics,''
  Nature Phys. 8, 393 (2012).
\bibitem{Casadio:2015rwa}
  R.~Casadio, O.~Micu, D.~Stojkovic,
  %``Inner horizon of the quantum Reissner-Nordstr?m black holes,''
  JHEP 1505, 096 (2015).

\bibitem{aliJHEP}
A. F. Ali, JHEP 1209, 067 (2012).
\bibitem{majumPLB}
B. Majumder, Phys. Lett. B 701, 384 (2011).
\bibitem{binaPRD}
A. Bina, S. Jalalzadeh, A. Moslehi, Phys. Rev. D 81, 023528 (2010).
\bibitem{chenNPPS}
P. Chen, R. J. Adler, Nucl. Phys. Proc. Suppl. 124, 103 (2003).
\bibitem{adlerGRG}
R.J. Adler, P. Chen, D.I. Santiago, Gen. Relativ. Gravit. 33, 2101 (2001).
\bibitem{xiangwenJHEP}
L. Xiang, X. Q.Wen, JHEP 0910, 046 (2009).
\bibitem{kimsonJHEP}
W. Kim, E. J. Son, M. Yoon, JHEP 0801, 035 (2008).
\bibitem{nozariEPL}
K. Nozari, S.H. Mehdipour, EPL 84, 20008 (2008).

\bibitem{majumGRG}
B. Majumder, Gen. Relativ. Gravit. 45, 2403 (2013).
\bibitem{chenahep}
D.Y. Chen, H.W. Wu, H.T. Yang, Adv. High Energy Phys. 432412 (2013).
\bibitem{chenjhep}
D.Y. Chen, Q.Q. Jiang, P.Wang, H.T. Yang, JHEP 11, 176 (2013).
\bibitem{chenjcap}
D.Y. Chen, H.W. Wu, H.T. Yang, JCAP 03, 036 (2014).
\bibitem{barguenoPLB}
P. Bargueno, E.C. Vagenas, Phys. Lett. B 742, 15 (2015).
\bibitem{mubenrong}
B.R. Mu, P.Wang, H.T. Yang, Adv. High Energy Phys. 898916 (2015).
\bibitem{pengwang}
P.~Wang, H.~Yang and S.~Ying, Int.\ J.\ Theor.\ Phys. 55, no. 5, 2633 (2016).
\bibitem{anaclePLB}
M.A. Anacleto, F.A. Brito, E. Passos, Phys.Lett.B 749, 181 (2015).
\bibitem{fengEPJC}
Z.~W.~Feng, H.~L.~Li, X.~T.~Zu, S.~Z.~Yang, Eur.\ Phys.\ J.\ C 76, no. 4, 212 (2016).

\bibitem{Kober:2010sj}
M.~Kober, Phys.\ Rev.\ D 82, 085017 (2010).

\bibitem{Umetsu1}
K. Umetsu, Int. J. Mod. Phys. A 25,4123 (2010).

\bibitem{Umetsu2}
K. Umetsu, Phys. Lett. B 692,61 (2010).

\bibitem{Chatrchyan:2012me}
S.~Chatrchyan {\it et al.} [CMS Collaboration], JHEP 1209, 094 (2012).
%  doi:10.1007/JHEP09(2012)094
% [arXiv:1206.5663 [hep-ex]].


% Please avoid comments such as "For a review'', "For some examples",
% "and references therein" or move them in the text. In general,
% please leave only references in the bibliography and move all
% accessory text in footnotes.

% Also, please have only one work for each \bibitem.


\end{thebibliography}
\end{document}